\documentclass[utf8, final]{FrontiersinHarvard} %
\usepackage[onehalfspacing]{setspace}
\usepackage[acronym]{glossaries}
\usepackage{url}
\usepackage{hyperref}
\usepackage{lineno}
\usepackage{microtype}
\usepackage{subcaption}
\usepackage{amsmath}
\usepackage{amssymb}
\usepackage{natbib}
\usepackage{makecell}

\newcommand{\fracPow}[3]{\left( \dfrac{#1}{#2} \right)^{#3}}

\def\keyFont{\fontsize{8}{11}\helveticabold }
\def\firstAuthorLast{Alves Batista \& Saveliev} 
\def\Authors{Rafael {Alves Batista}\,$^{1,2,*}$, Andrey Saveliev\,$^{3,4,*}$}

\def\Address{$^{1}$Sorbonne Universit{\'e}, Institut d'Astrophysique de Paris (IAP), CNRS UMR 7095\\ 98 bis bd Arago 75014, Paris, France \\
$^{2}$Sorbonne Universit{\'e}, Université Paris Cité, Lab. de Physique Nucl{\'e}aire et des Hautes {\'E}nergies (LPNHE), CNRS/IN2P3, 4 place Jussieu, F-75005, Paris, France \\
$^{3}$Immanuel Kant Baltic Federal University, Institute of High Technology, Ul. A. Nevskogo 14, 236016 Kaliningrad, Russia \\
$^{4}$Lomonosov Moscow State University, Faculty of Computational Mathematics and Cybernetics, GSP-1, Leninskiye Gory 1-52, 119234 Moscow, Russia
}
\def\corrAuthor{Rafael Alves Batista, Andrey Saveliev}
\def\corrEmail{rafael.alves\_batista@iap.fr, anvsavelev@kantiana.ru}

\begin{document}
\newacronym[\glslongpluralkey = {active galactic nuclei}]{AGN}{AGN}{active galactic nucleus}
\newacronym{CMB}{CMB}{cosmic microwave background}
\newacronym{CR}{CR}{cosmic ray}
\newacronym{CRB}{CRB}{cosmic radio background}
\newacronym{DGRB}{DGRB}{diffuse gamma-ray background}
\newacronym{EBL}{EBL}{extragalactic background light}
\newacronym{EGMF}{EGMF}{extragalactic magnetic field}
\newacronym{EM}{EM}{electromagnetic}
\newacronym{HE}{HE}{high energy}
\newacronym{IACT}{IACT}{imaging atmospheric Cherenkov telescope}
\newacronym{ICS}{ICS}{inverse Compton scattering}
\newacronym{IGM}{IGM}{intergalactic medium}
\newacronym{IGMF}{IGMF}{intergalactic magnetic field}
\newacronym{MHD}{MHD}{magnetohydrodynamics}
\newacronym{PIC}{PIC}{particle-in-cell}
\newacronym{RNG}{RNG}{random number generator}
\onecolumn
\firstpage{1}

\title[Simulations of Electromagnetic Cascades with Plasma Instabilities]{Simulations of Electromagnetic Cascades in the Intergalactic Medium with Plasma Instabilities: the \texttt{grplinst} Code}

\author[\firstAuthorLast ]{\Authors} 
\address{} 
\correspondance{} 
\extraAuth{}


\maketitle

\begin{abstract}

Electromagnetic cascades initiated by TeV gamma rays from distant blazars provide one of the cleanest indirect probes of the intergalactic medium and, in particular, of intergalactic magnetic fields. However, their interpretation is not completely clear because of an open theoretical question: the extent to which the electron-positron beams generated in the cascade can lose energy through collective plasma processes. If this occurs before inverse Compton scattering produces secondary gamma rays, then the cascade is quenched. Here we present \texttt{grplinst}, a plugin for the CRPropa framework that models plasma-instability cooling acting on electrons and positrons during propagation. We describe the implementation of several prescriptions proposed in the literature, and present illustrative examples. The code is suitable both for bracketing theoretical uncertainties and for performing systematic studies of how plasma-instability assumptions propagate into gamma-ray observables and inferred intergalactic magnetic-field constraints, which is essential for interpreting current and forthcoming observations by high-energy gamma-ray observatories.

\tiny
\keyFont{ \section{Keywords:} electromagnetic cascades, gamma-ray astronomy, blazars, intergalactic medium, plasma instabilities, intergalactic magnetic fields, Monte Carlo methods, CRPropa}

\end{abstract}

\section{Introduction} 
\label{sec:intro}

High-energy gamma rays emitted by distant astrophysical objects propagate in the \gls{IGM} before reaching Earth. Depending on their energy and the distance travelled, they can interact with background photons producing electron-positron pairs. In the TeV energy band, the predominant target photon field is the \gls{EBL}~\citep{gould1967a}. The pairs can subsequently undergo further interactions, namely inverse Compton scattering, chiefly with the \gls{CMB}, generating energetic photons which can, again, undergo interactions, ultimately producing an electromagnetic cascade. The resulting cascade radiation in the GeV range has long been recognised as a sensitive probe of the physics of intergalactic space, in particular of \glspl{IGMF}~\citep{plaga1995a, neronov2009a, neronov2010a, tavecchio2010a, alvesbatista2021a}.

The exact contribution of secondary (cascade) photons depends strongly on the intrinsic spectrum of the source. Moreover, the ill-understood \glspl{IGMF} can significantly suppress the observed cascade by deflecting the pairs away from the line of sight, which can be particularly relevant for beamed sources. If plasma instabilities indeed act as an additional cooling term for electrons and positrons, this might dominate over inverse Compton scattering, ultimately preventing the production of secondary photons. As a consequence, \gls{IGMF} constraints based on gamma-ray observations would be significantly compromised, which can be problematic given that this is arguably the most promising strategy for probing cosmic magnetism over large scales.

An uncertainty in this picture is whether the pair beam produced by the cascade behaves as a collection of independent particles. Collective interactions between the lepton pairs and the tenuous background plasma that composes the \gls{IGM} could lead to the growth of plasma instabilities, which could drain energy from the beam before the electrons can be reprocessed into secondary photons~\citep{broderick2012a,schlickeiser2012a, miniati2013a, schlickeiser2013a, sironi2014a, supsar2014a, chang2014a, chang2016a, kempf2016a, rafighi2017a, vafin2018a, shalaby2018a, vafin2019a, alvesbatista2019g, shalaby2020a, perry2021a, alawashra2022a, alawashra2024a, alawashra2025b}. 

Many studies have argued that beam-plasma instabilities may grow rapidly enough to quench the cascade and convert beam energy into heat in the \gls{IGM}~\citep{broderick2012a, chang2012a}. Other studies have found that non-linear effects, finite angular spread, background inhomogeneities, and magnetic broadening can strongly reduce the effective energy transfer, even when unstable modes are present~\citep{sironi2014a}. Therefore, until this issue is settled, it is important to be able to explore how different assumptions about plasma instabilities would affect gamma-ray observations.

The archetypal class of objects for this type of study are blazars, as this type of \gls{AGN} is characterised by tightly collimated jets approximately pointing toward Earth. Hard-spectrum blazars such as 1ES~0229+200 have been the primary targets~\citep{bonnoli2015a, chang2019a, costamante2020a, magic2023a}. The pairs produced along the way are also highly collimated due to the beam, increasing the effective density of charged particles that could spawn instabilities.

Another interesting consequence of plasma instabilities is their potential effect on the thermal history of the universe. In particular, the energy lost by the electron-positron pairs could, in principle, be dumped into the \gls{IGM}, heating it up~\citep{chang2012a, puchwein2012a, broderick2014a, lamberts2015a}. Nevertheless, there is no guarantee that the energy lost by the pairs is efficiently transferred to the background plasma, and even if this is the case, it is not clear how much of this energy would be converted into heat. It is possible that the instabilities will simply spread the beam, without significantly heating the \gls{IGM}~\citep{sironi2014a, perry2021a, alawashra2024a}.

The main difficulty in assessing the role of the instabilities in this context stems from the difficulty in performing simulations of such low-density environments. In particular, the pair beam can be up to $10^{24}$ times less dense than the surrounding plasma~\citep{schlickeiser2012a}, leading to a highly non-linear interaction regime that cannot be easily captured by computational methods such as \gls{PIC} simulations (see, e.g., \citet{pohl2020a} for a detailed discussion).

\bigskip
Considering the above, it is clear that the question of whether plasma instabilities can quench electromagnetic cascades is still open. Until this type of phenomenon can be accurately modelled in simulations and~/~or constrained by observations, it is important to be able to explore how different assumptions about plasma instabilities would affect gamma-ray observations. A possible phenomenological path explored here is to implement effective cooling terms in cascade simulations, which can be used to bracket the range of possible effects of plasma instabilities on cascade development and gamma-ray observations. 

This is the motivation behind the development of \texttt{grplinst}, a plugin for the CRPropa framework~\citep{alvesbatista2016a, alvesbatista2022a} that models plasma-instability cooling as an effective continuous energy-loss term acting on electrons and positrons during cascade propagation. The code is designed to be modular, allowing users to vary density, temperature, beam geometry, and cooling model independently, and to extend the code from either C++ or Python.

The paper is organised as follows. First, we provide a description of plasma instabilities and their effect on electromagnetic cascades in section~\ref{sec:instabilities}. Section \ref{sec:code} contains a description of the software architecture, followed by some example applications in section~\ref{sec:examples} and a general discussion in section~\ref{sec:discussion}. Finally, in section~\ref{sec:outlook}, we summarise our findings and discuss future prospects.

\section{Plasma Instabilities in the IGM}
\label{sec:instabilities}

If the density of electron--positron pairs in a given medium is sufficiently high, collective effects may become important. A useful characteristic scale for assessing such effects is the plasma wavelength ($\lambda_\text{pl}$), associated with the electron plasma frequency through
\begin{equation}
    \lambda_\text{pl} \equiv \dfrac{2 \pi c}{\omega_\text{pl}} \,,
    \label{eq:plasmaWavelength}
\end{equation}
where the plasma frequency is defined as
\begin{equation}
    \omega_\text{pl} = \sqrt{\dfrac{e^{2} n_\text{IGM}}{\epsilon_{0} m_{e}}} \,.
\end{equation}
Here, $e$ is the elementary charge, $n_\text{IGM}$ is the number density of the electrons in the background plasma, $\epsilon_0$ is the vacuum permittivity, and $m_e$ is the electron mass. This wavelength is related to the plasma skin depth ($\delta_\text{pl}$) through
\begin{equation}
    \delta_\text{pl} \equiv \dfrac{c}{\omega_\text{pl}} = \dfrac{\lambda_\text{pl}}{2 \pi} \,.
    \label{eq:skinDepth}
\end{equation}

The wavelength and growth rate of the fastest-growing beam--plasma mode are not fixed by this characteristic scale alone, but depend on the beam-to-plasma density ratio, the beam energy and angular distributions, and the properties of the background plasma.

For the densities involved in blazar-induced pair beams interacting with the \gls{IGM}, plasma effects may be relevant for the evolution of the beam. In this section, we present some of the instabilities that might arise from interactions between the beam and the background plasma. We focus on the modes that have been proposed as relevant for ultrarelativistic pair beams propagating through the \gls{IGM}.

The medium with which the beam interacts is the \gls{IGM}, whose physical number density evolves with redshift ($z$) as
\begin{equation}
    n_\text{IGM} = n_{\text{IGM},0} \, \left( 1 + z \right)^{3} \,.
\end{equation}
We adopt a fiducial present-day density of $n_{\text{IGM},0} = 0.1 \; \text{m}^{-3}$ and a temperature of $T_{\text{IGM},0} = 10^{4} \; \text{K}$~\citep{meiksin2009a, mcquinn2016a}. These quantities can vary by orders of magnitude with environment; voids are less dense and colder than galaxy clusters. Since blazar-induced cascades predominantly develop in low-density intergalactic environments, these values are representative of the conditions considered here.

Electron--positron pairs are produced by pair production. Their density depends on the intrinsic source properties, pair-production rate, distance from the source, beam geometry, and cooling history. TeV-emitting blazars have typical isotropic-equivalent luminosities in the range $10^{37} \; \text{W}$ to $10^{40} \; \text{W}$~\citep{celotti2008a, sikora2016a, chen2018a, sol2022a}. In the fiducial cascade configuration considered here, we adopt a characteristic pair density of $n_\text{beam} \sim 10^{-16} \; \text{m}^{-3}$.

\subsection{Models} 
\label{sec:models}

To investigate how plasma instabilities develop, one must consider two characteristic timescales. The first is the instability growth time, $\mathcal{T}_{i}$, which measures how rapidly collective plasma modes develop. The second is the energy-loss time of cascade electrons and positrons ($\tau_{i}$), which quantifies how efficiently they transfer energy to the medium. While both are informative, the latter is the relevant input for spectral calculations. Since a rigorous determination of $\tau_{i}$ is often not available, we conservatively set $\tau_{i} = \mathcal{T}_{i}$.
In reality,  $\tau_{i}$ may exceed $\mathcal{T}_{i}$ by orders of magnitude~\citep{grognard1975a,pavan2011a}. This prescription yields the most robust lower bound on the cascade photon flux.

Although many types of plasma instabilities can in principle co-operate, in practice each model identifies a single dominant channel.

The variety of approaches found in the literature differs not only in their choice of instability, but also in methodology: some rely on analytic calculations, whereas others perform numerical \gls{PIC} simulations attempting to capture the non-linear effects. They are summarised in table~\ref{tab:models}.
\begin{table}[hbt!]
    \centering
    \caption{Primary unstable modes or subsequent non-linear mechanisms considered in the different models, together with the treatment employed in the original works.}
    \begin{tabular}{ll}
        \hline
        \hline
        \textbf{model} & \textbf{mode or mechanism and treatment} \\
        \hline
        \citet{bret2010a} &
        \makecell[l]{
            two-stream, filamentation, and oblique modes\\
            cold-fluid and kinetic linear theory; PIC results
        } \\[10pt]

        \citet{broderick2012a} &
        \makecell[l]{
            oblique instability in the reactive and kinetic regimes\\
            analytical growth-rate and cooling-time estimates
        } \\[10pt]

        \citet{miniati2013a} &
        \makecell[l]{
            kinetic electrostatic instability; non-linear Landau damping\\
            Monte Carlo cascade modelling and kinetic theory
        } \\[10pt]

        \citet{schlickeiser2012a} &
        \makecell[l]{
            oblique electrostatic instability; subsequent modulation\\
            analytical linear and non-linear theory
        } \\[10pt]

        \citet{sironi2014a} &
        \makecell[l]{
            oblique and quasi-longitudinal modes\\
            two- and three-dimensional PIC simulations
        } \\[10pt]

        \citet{vafin2018a} &
        \makecell[l]{
            electrostatic instability; fastest growth for quasi-parallel modes\\
            linear analysis and PIC simulations
        } \\[10pt]

        \citet{shalaby2020a} &
        \makecell[l]{
            longitudinal beam--plasma instability in an inhomogeneous medium\\
            relativistic kinetic theory and PIC validation
        } \\
        \hline
        \hline
    \end{tabular}
    \label{tab:models}
\end{table}

\subsubsection{Models by \citet{bret2010a}}

The fastest-growing modes of two types of instabilities are considered by \citet{bret2010a}: the filamentation and the two-stream instabilities. Their characteristic timescales are given, respectively, by
\begin{equation}
	\tau(E_e) = 2.5 \times 10^{9} \; \text{s} \; \fracPow{E_e}{1 \; \text{TeV}}{\frac{1}{2}} \, \fracPow{n_\text{beam}}{10^{-16} \; \text{m}^{-3}}{-\frac{1}{2}} \,,
\end{equation}
and 
\begin{equation}
	\tau(E_e) = 1.6 \times 10^{9} \; \text{s} \; \fracPow{E_e}{1 \; \text{TeV}}{} \, \fracPow{n_\text{beam}}{10^{-16} \; \text{m}^{-3}}{-\frac{1}{3}} \, \fracPow{n_\text{IGM}}{0.1 \; \text{m}^{-3}}{-\frac{1}{6}}  \,.
\end{equation}

Note that these expressions are derived for highly idealised beam--plasma configurations; they are not completely comparable with the other models discussed here.
They are included here as phenomenological linear-growth reference prescriptions, rather than as quantitative models for realistic blazar-induced pair beams, which are extremely dilute, charge neutral, and have a finite angular spread.

\subsubsection{Model by \citet{broderick2012a}} 

Two regimes are defined, `cold' and `warm', depending  on whether the value of the beam density ($n_\text{beam}$) is below or above a critical value $n_\text{crit}$, given by
\begin{equation}
	n_\text{crit} = 1.6 \times 10^{-13} \fracPow{E_e}{1 \; \text{TeV}}{-2} \, \fracPow{n_\text{IGM}}{0.1 \; \text{m}^{-3}}{} \; \text{m}^{-3} \,.
\end{equation}
In both cases, the oblique instability dominates. The characteristic timescale is given by
\begin{equation}
	\tau(E_e) \simeq 
	\begin{cases}
		7 \times 10^{7} \; \text{s} \; \fracPow{E_e}{1 \; \text{TeV}}{-1} \fracPow{n_\text{beam}}{10^{-16} \; \text{m}^{-3}}{-1} \fracPow{n_\text{IGM}}{0.1 \; \text{m}^{-3}}{\tfrac{1}{2}}   & \text{if} \;\;\;  n_\text{beam} < n_\text{crit} \, \\[12pt]
		5 \times 10^{5} \; \text{s} \; \fracPow{E_e}{1 \; \text{TeV}}{1/3} \fracPow{n_\text{beam}}{10^{-16} \; \text{m}^{-3}}{-\tfrac{1}{3}} \fracPow{n_\text{IGM}}{0.1 \; \text{m}^{-3}}{-\tfrac{1}{6}}  & \text{if} \;\;\;  n_\text{beam} \geq n_\text{crit} \,.
	\end{cases}
	\label{eq:tauBroderick}
\end{equation}

\subsubsection{Model by \citet{miniati2013a}}

In this model, Langmuir waves and non-linear Landau damping are the key processes to consider. In their treatment, the finite spread of the transverse momentum component of the beam, obtained from Monte Carlo simulations of cascade development, was explicitly included. Their analysis shows that plasma oscillations contribute negligibly to the energy losses of electron-positron beams, since inhomogeneities in the \gls{IGM} disrupt the resonance between Langmuir modes and the beam. The corresponding energy-loss time, $\tau$, was derived by combining Monte Carlo simulations (to characterise the evolving beam properties) with analytic estimates of the instability growth rate. It can be written as
\begin{equation}
	\tau(E_e) \simeq \tau_\text{IC} \, \mathfrak{T} (D) \, (1 + z)^{2} 
	= 3.9 \times 10^{13} \, \text{s} \, (1 + z)^{-2} 
	\, \fracPow{E_e}{1 \; \text{TeV}}{-1} \, \mathfrak{T}_\text{B}(D) \,,
	\label{eq:tauMiniati}
\end{equation}
where $E_{e}$ is the electron (or positron) energy, $D$ is the co-moving distance to the source, and $\mathfrak{T}_\text{B}(D)$ is a value that depends on the distance\footnote{The exact values are given in table~2 of \citet{alvesbatista2019g}.}, ranging from $\sim 3.5$ for small $D$ to $0.75$ for $D \simeq 1000 \; \text{Mpc}$. This reflects the predominance of non-linear Landau damping at small distances from the blazar, whilst \gls{IGM} inhomogeneities become increasingly relevant farther out, ultimately stabilising the beam and rendering plasma instabilities subdominant in this model.

\subsubsection{Model by \citet{schlickeiser2012a}}

In this model, different plasma effects suppress electromagnetic cascades in different regimes. When the beam density is above a critical threshold $n_\text{crit}$ (the \emph{strong blazar} regime), the modulation instability dominates. Conversely, for $n_\text{beam} < n_\text{crit}$ (the \emph{weak blazar} regime), the modulation instability fails to develop, and non-linear Landau damping becomes the main channel, depositing energy into electrostatic and electromagnetic fluctuations of the background plasma. The corresponding energy-loss time is
\begin{equation}
	\tau(E_e) \simeq 
	\begin{cases}
		\begin{split}
			5 \times 10^{14} \; \text{s} \; \fracPow{E_e}{1 \; \text{TeV}}{\tfrac{5}{3}} \, \fracPow{n_\text{beam}}{10^{-16} \; \text{m}^{-3}}{\tfrac{1}{3}} \, \fracPow{n_\text{IGM}}{0.1 \; \text{m}^{-3}}{-\tfrac{5}{6}} \, \fracPow{T_\text{IGM}}{10^4 \; \text{K}}{-2} & \text{if} \;\;\;  n_\text{beam} \le n_\text{crit} \,, \\[12pt]
			8 \times 10^{6} \; \text{s} \; \fracPow{E_e}{1 \; \text{TeV}}{\tfrac{1}{3}} \, \fracPow{n_\text{beam}}{10^{-16} \; \text{m}^{-3}}{-\tfrac{1}{3}} \, \fracPow{n_\text{IGM}}{0.1 \; \text{m}^{-3}}{-\tfrac{1}{6}} \, \mathfrak{T} (n_\text{IGM},T_\text{IGM}) & \text{if} \;\;\;  n_\text{beam} > n_\text{crit} \,,
		\end{split}
	\end{cases}
	\label{eq:tauSchlickeiser}
\end{equation}
with
\begin{equation}
	\mathfrak{T} (n_\text{IGM},T_\text{IGM}) = 1 + \frac{5}{4} \ln\!\left( \frac{T_\text{IGM}}{10^{4} \; \text{K}} \right) - \frac{1}{4} \ln\fracPow{n_\text{IGM}}{0.1 \; \text{m}^{-3}}{} \,,
\end{equation}
and the critical beam density given by
\begin{equation}
	n_\text{crit} = 2.5 \times 10^{-19} \,
	\fracPow{E_e}{1 \; \text{TeV}}{-1} \fracPow{n_\text{IGM}}{0.1 \; \text{m}^{-3}}{} \fracPow{T}{10^{4} \; \text{K}}{2}{} \; \text{m}^{-3}   \,.
\end{equation}

An important feature of this model is the temperature dependence. In the weak blazar regime, the cooling rate scales quadratically with $T_\text{IGM}$, while in the strong blazar case this dependence becomes much weaker. This contrast arises from the onset of the modulation instability, which is jointly controlled by the beam density $n_\text{beam}$ and the IGM temperature $T_\text{IGM}$.

\subsubsection{Model by \citet{sironi2014a}}

This model is based on a model similar to the one by \citet{broderick2012a} described above. They differ by some numerical factors in the equations, but most importantly, their overall behaviours are qualitatively different from each other. Within the framework of this model, the results from \gls{PIC} simulations suggest that less than 10\% of the beam energy is transferred to the background plasma, in contrast to the $\sim 50\%$ estimated by \citet{broderick2012a}. This discrepancy arises, in part, from the treatment of the transverse momentum component, which is inherited from the momentum distribution of the pairs generated in the cascade.

The authors distinguish two cases, the \emph{cold-plasma beam} and the \emph{warm-plasma beam}, depending on whether the beam plasma density ($n_\text{beam}$) is above or below the value $n_\text{crit}$, respectively. For both cases they find that the oblique instability is the most relevant one, resulting in the energy-loss time
\begin{equation}
\tau(E_e) \simeq 
	\begin{cases}
		1.4 \times 10^{7} \; \text{s} \; \fracPow{E_e}{1 \; \text{TeV}}{-1} \fracPow{n_\text{beam}}{10^{-16} \; \text{m}^{-3}}{-1} \fracPow{n_\text{IGM}}{0.1 \; \text{m}^{-3}}{\tfrac{1}{2}}   & \text{if} \;\;\;  n_\text{beam} < n_\text{crit} \, \\[12pt]
		9.6 \times 10^{5} \; \text{s} \; \fracPow{E_e}{1 \; \text{TeV}}{\tfrac{1}{3}} \fracPow{n_\text{beam}}{10^{-16} \; \text{m}^{-3}}{-\tfrac{1}{3}} \fracPow{n_\text{IGM}}{0.1 \; \text{m}^{-3}}{-\tfrac{1}{6}}  & \text{if} \;\;\;  n_\text{beam} \geq n_\text{crit} \,,
	\end{cases}
	\label{eq:gamma:plasma:tauSironi}
\end{equation}
where $n_\text{crit}$ is given by
\begin{equation}
	n_\text{crit} = 8 \times 10^{-14}  \, \left( \dfrac{E_{e}}{1 \; \text{TeV}} \right)^{-2} \left( \dfrac{n_\text{IGM}}{0.1 \; \text{m}^{-3}} \right) \; \text{m}^{-3} \,.
\end{equation}

\subsubsection{Model by \citet{vafin2019a}}

In this model, based on \gls{PIC} simulations, the modulation instability is the dominant channel for beam energy losses. The characteristic energy-loss timescale
\begin{equation}
	\tau(E_e) \simeq 1.9 \times 10^{11} \; \text{s} 
	\fracPow{E_e}{1 \; \text{TeV}}{\tfrac{4}{3}} \fracPow{n_\text{beam}}{10^{-16} \; \text{m}^{-3}}{-\tfrac{1}{3}}  \fracPow{n_\text{IGM}}{0.1 \; \text{m}^{-3}}{\tfrac{1}{3}} \fracPow{T_\text{IGM}}{10^{4} \; \text{K}}{-1}
	\label{eq:gamma:plasma:tauVafin}
\end{equation}
In their simulations, non-linear Landau damping is explicitly resolved. The instability peaks at frequencies around $\sim 10^{-5} \; \omega_\text{p}$, notably lower than the $\sim 10^{-3}\omega_\text{p}$ expected from linear growth analyses.  The stabilisation occurs once the energy density stored in plasma waves reaches a fixed fraction of the beam energy, effectively quenching the electrostatic mode. This criterion implicitly defines an efficiency factor linking the beam energy to plasma heating. In line with the other models considered here, we adopt the maximal efficiency, corresponding to the scenario of strongest cascade suppression.

\subsubsection{Model by \citet{shalaby2020a}}

The authors investigate the longitudinal stability in the presence of a density inhomogeneity in the background plasma. They find that the energy-loss time is given by (eq. 4.16 of their paper)
\begin{equation}
	\tau(E_e) = 3.2 \times 10^{12} \; \text{s} \; \fracPow{E_e}{1 \; \text{TeV}}{\frac{6}{5}} \, \fracPow{n_\text{beam}}{10^{-16} \; \text{m}^{-3}}{-\frac{2}{5}} \, \fracPow{n_\text{IGM}}{0.1 \; \text{m}^{-3}}{-\frac{1}{10}}   \,.
\end{equation}

\section{Code design}
\label{sec:code}

The \texttt{grplinst} package is an external plugin for the CRPropa Monte Carlo framework~\citep{alvesbatista2016a, alvesbatista2022a} that adds an effective cooling term for electrons and positrons. It is intended to emulate energy losses associated with beam-plasma instabilities in blazar-induced pair beams propagating through the \gls{IGM}. The original public implementation and validation were presented by ~\citet{alvesbatista2019g}, where several widely cited prescriptions from the literature were compared and used as bounding cases for cascade quenching. 

The version presented here, \texttt{grplinst v2}, modernises the code architecture by decoupling the beam properties, the background medium, and the plasma instability prescription, offering more flexibility for studying how each parameter that goes into the model affects the predictions. Moreover, this modularity allows for extensions of the models, aiming to make them more realistic.

The code is written in C++11, with optional Python bindings generated via SWIG~\citep{beazley1996a} and NumPy~\citep{vanderwalt2011a, harris2020a} typemaps for interoperability with efficient arrays.

The design is compositional, with abstract objects acting as building blocks for simulation-building:
\begin{itemize}[noitemsep, nolistsep]
    \item \texttt{MediumDensity}: describes the properties of the medium wherein the beam propagates (e.g., the \gls{IGM});
    \item \texttt{MediumTemperature}: the temperature of the medium, which is relevant for some instability prescriptions;
    \item \texttt{Flow}: an abstract base providing describing the pair beam; a particular sub-class is \texttt{FlowJet1D}, which reads tabulated profiles (distance and density profile) from files and interpolates along a jet axis, with a finite ``emission geometry'' controlling the volume used for density normalisation;
    \item \texttt{PlasmaInstability}: this is the main class that computes the energy loss time $\tau$ for a given set of beam and medium properties, according to a specific instability prescription.
\end{itemize}
These building blocks are extendable not only from C++ but also from Python, through SWIG's ``directors'' for the abstract classes, allowing users to implement custom models without modifying the core codebase.

The \texttt{PlasmaInstability} module derives from CRPropa's \texttt{Module} class with a \texttt{process} function that acts on electrons and positrons, applying energy losses according to the specified instability model. 

The plugin essentially adds an energy-loss term affecting electrons and positrons, approximated as a continuous process. The energy lost at each propagation step is estimated as
\begin{equation}
	-\dfrac{\text{d}E_e}{\text{d}x} \left( E_e, \vec{x}, z \right) = \eta \dfrac{E_e}{c \, \tau(E_e, \vec{x}, z)} \,,
\end{equation}
where $\tau$ is the cooling time of the instability, given by the equations in section~\ref{sec:models}, and $\eta$ is an efficiency parameter that rescales the effective coupling between the beam and the phenomenological cooling prescription. The case $\eta = 1$ corresponds to the most extreme assumption, namely that the effective plasma-cooling time is as short as allowed by the adopted model. Smaller values of $\eta$ provide a simple way to emulate incomplete dissipation or saturation effects without modifying the functional dependence of the prescription itself.

\begin{figure}[htb!]
	\centering
	\includegraphics[width=0.8\columnwidth]{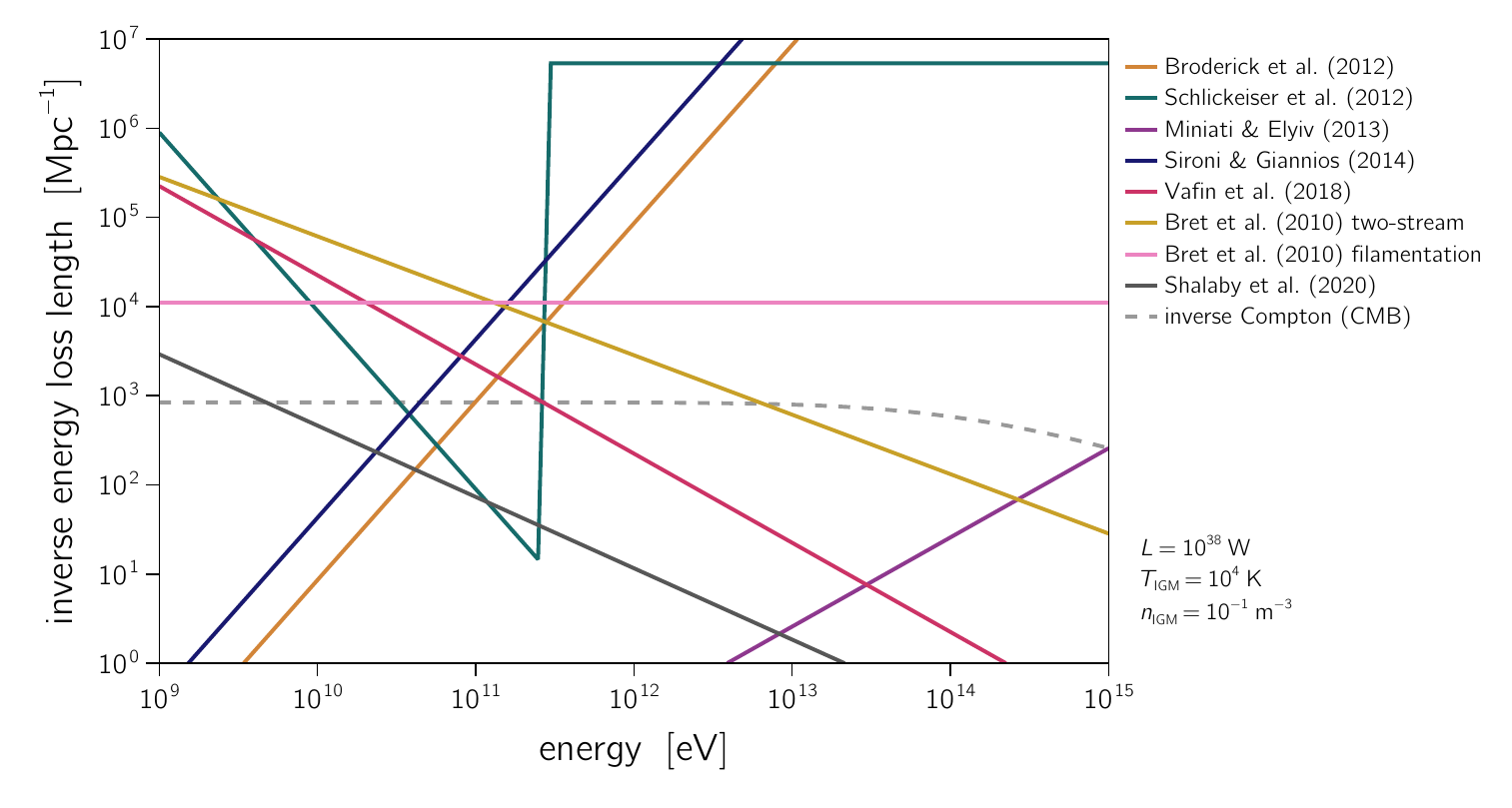}
	\caption{Cooling rates for the different plasma instability models considered here, compared to the mean free path for inverse-Compton scattering with the \gls{CMB} (dashed line). The parameters used are $n_\text{IGM} = 0.1 \; \text{m}^{-3}$, $T_\text{IGM} = 10^4 \; \text{K}$, and $n_\text{beam} = 10^{-16} \; \text{m}^{-3}$.}
	\label{fig:energyLoss}
\end{figure}

Figure~\ref{fig:energyLoss} summarises the cooling scales associated with the different prescriptions implemented in the code and compares them with inverse-Compton losses on the \gls{CMB}.

Operationally, this implementation should be interpreted as a phenomenological layer on top of the standard machinery of CRPropa. It does not solve the kinetic plasma problem self-consistently, nor does it evolve wave spectra or distribution functions. Instead, it maps a chosen instability model onto a local continuous loss term that is sampled during particle propagation.

\section{Examples}
\label{sec:examples}

\begin{figure}[htb!]
	\centering
	\includegraphics[width=0.8\columnwidth]{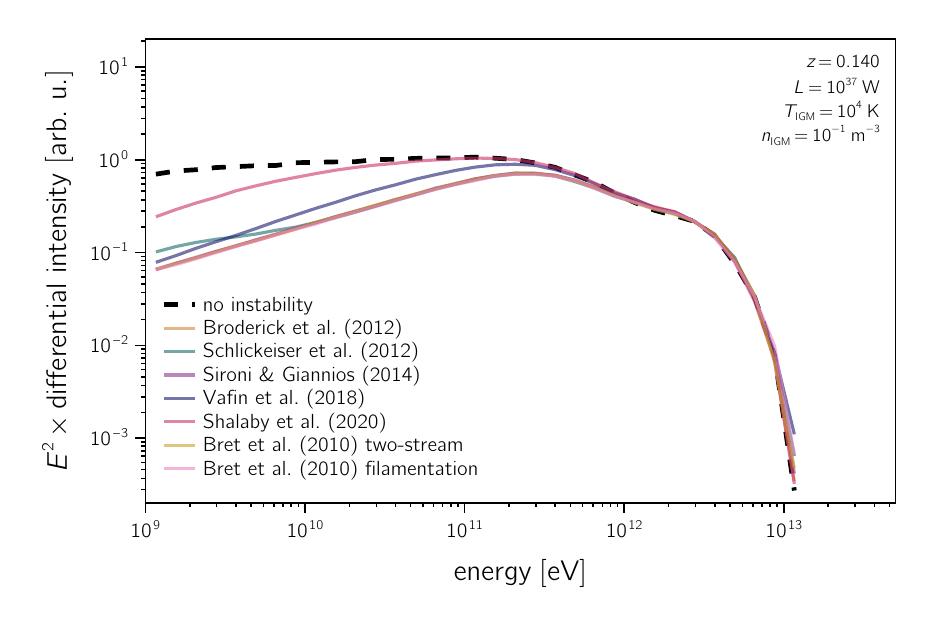}
	\caption{Simulated gamma-ray fluxes from the blazar 1ES~0229+200 ($z=0.14$). The case without plasma instabilities is shown as a dashed line, whereas solid coloured lines represent the different plasma instability prescriptions. The parameters used are shown at the top right of the figure. The intrinsic source spectrum is a power law with index $\alpha = 1.5$ and an exponential cutoff at $E_\text{cut} = 8 \; \text{TeV}$.}
	\label{fig:spectra}
\end{figure}

The main use case of \texttt{grplinst} is to propagate theoretical uncertainty in plasma-instability cooling into observable quantities. The simplest possible application is to run one-dimensional cascade calculations for a given set of source and medium parameters, comparing the resulting gamma-ray spectra with and without plasma-instability losses, as done by \citet{alvesbatista2019g}, which also provides a first demonstration of the code. Figure~\ref{fig:energyLoss} provides one such benchmark by comparing the effective cooling lengths of the available models for a fiducial void environment. A particular application of this type is shown in figure~\ref{fig:spectra}, where the resulting spectra for the extreme blazar 1ES~0229+200, at $z=0.14$, are shown for the different models, compared to the case without plasma instabilities. The source is assumed to emit a power-law spectrum with index $\alpha = 1.5$ and an exponential cutoff at $E_\text{cut} = 8 \; \text{TeV}$, with an isotropic-equivalent luminosity of $L = 10^{37} \; \text{W}$.

\begin{figure}[htb!]
    \centering
    \includegraphics[width=0.8\columnwidth]{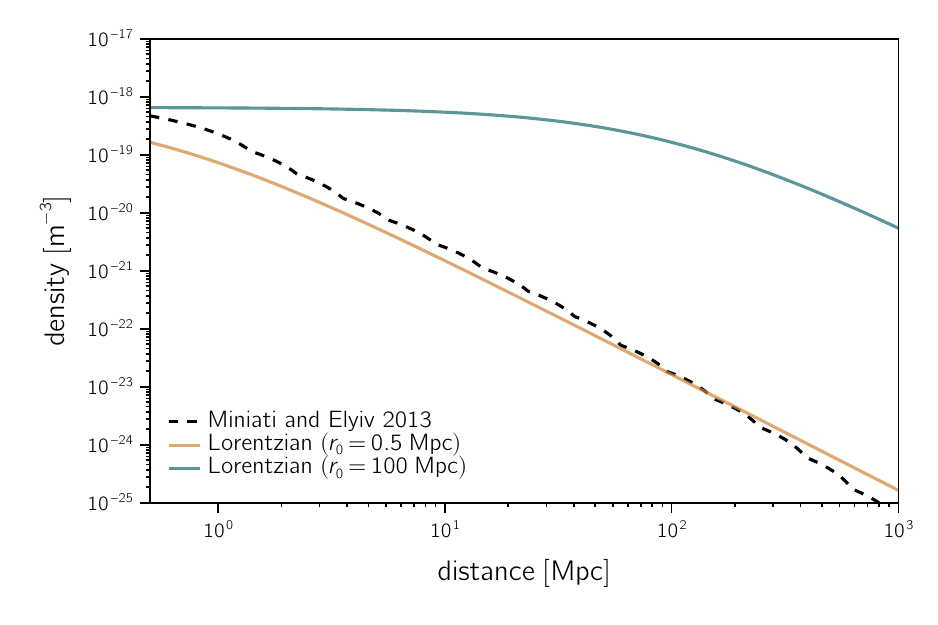}
    \caption{Beam number density as a function of distance from the source. The dashed line corresponds to the profile obtained by interpolating the results of \citet{miniati2013a} and rescaling them to the adopted luminosity. The solid lines show the two illustrative Lorentzian profiles defined by eq.~\ref{eq:lorentzian}, with characteristic distances $r_0 = 0.5 \; \text{Mpc}$ (orange) and $r_0 = 100 \; \text{Mpc}$ (green).}
    \label{fig:beamProfiles}
\end{figure}

In this version of \texttt{grplinst}, it is also possible to specify the distance profile of the beam, as discussed in section~\ref{sec:code}. To illustrate this option, we consider the energy-loss prescription of \citet{schlickeiser2012a}. However, instead of adopting the homogeneous beam profile used in figure~\ref{fig:spectra}, we consider three distance-dependent profiles, shown in figure~\ref{fig:beamProfiles}. The first is obtained by interpolating the beam densities calculated by \citet{miniati2013a} and rescaling them to the luminosity. The other two are described by Lorentzian profiles of the form
\begin{equation}
    n_\text{beam}(r) = \dfrac{n_{\text{beam},0}}{1 + \left( \dfrac{r}{r_0} \right)^2} \,,
    \label{eq:lorentzian}
\end{equation}
where $r$ is the distance from the source along the beam, $n_{\text{beam},0}(L)$ is the beam-density normalisation close to the source for a luminosity $L$, and $r_0$ is a characteristic distance. For the two illustrative profiles, we take $r_0 = 0.5 \; \text{Mpc}$ and $r_0 = 100 \; \text{Mpc}$.

\begin{figure}[htb!]
    \centering
    \includegraphics[width=0.8\columnwidth]{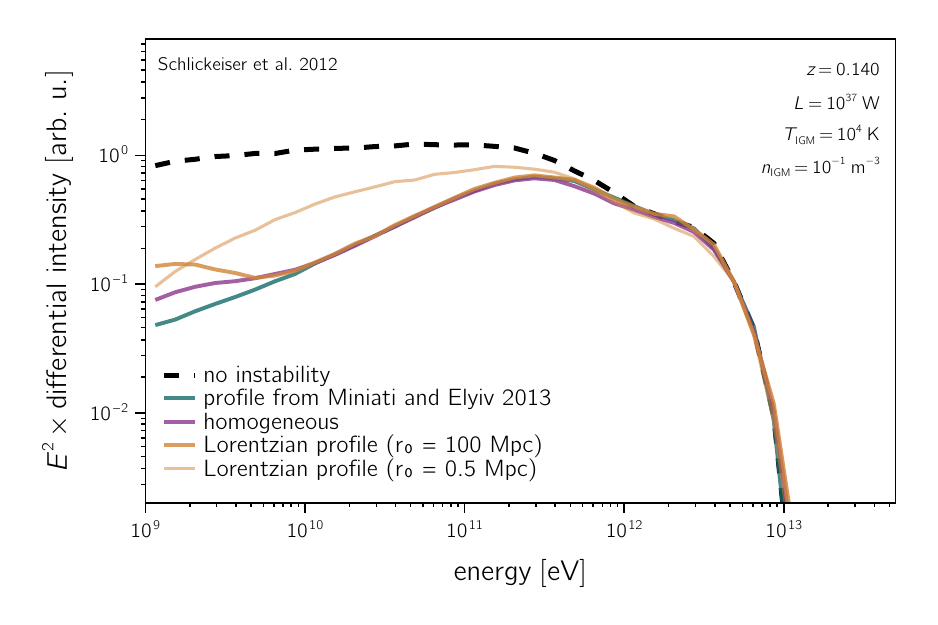}
    \caption{Simulated gamma-ray fluxes from the blazar 1ES~0229+200 ($z = 0.14$). The case without plasma instabilities is shown by the thick dashed line, whereas the solid coloured lines show the results obtained using the plasma-instability energy-loss prescription of \citet{schlickeiser2012a} and the beam-density profiles from figure~\ref{fig:beamProfiles}. The parameters used in the simulations are shown in the upper-right corner. The intrinsic source spectrum follows a power law with index $\alpha = 1.5$ and an exponential cutoff at $E_\text{cut} = 8 \; \text{TeV}$.}
    \label{fig:spectraBeam}
\end{figure}

Using the profiles shown in figure~\ref{fig:beamProfiles}, we simulate the development of the cascade with \texttt{grplinst} for a beam luminosity of $L = 10^{37} \; \text{W}$. The resulting spectra are shown in figure~\ref{fig:spectraBeam}.

The differences between the spectra in figure~\ref{fig:spectraBeam} arise due to the fact that the local beam density enters the plasma energy-loss time and effectively controls the accumulated cooling of the electrons and positrons along their trajectories.

Profiles for which the energy-loss rates are larger over the region wherein most pairs are produced lead to a larger integrated plasma energy loss and, consequently, to a stronger suppression of the secondary gamma-ray flux. The Lorentzian profile with $r_0 = 0.5 \; \text{Mpc}$ supports this argument: it decreases rapidly outside the immediate vicinity of the source (cf.~figure~\ref{fig:beamProfiles}), implying that pairs produced farther away from the source do not experience substantial plasma cooling. The profile with $r_0 = 100 \; \text{Mpc}$, in contrast, maintains a substantial beam density over a much larger fraction of the region where pair production is expected to take place, thereby producing a stronger modification of the cascade.

Note that the beam profiles used here are for illustrative purposes only. A self-consistent treatment of plasma instabilities would instead need to track the density of pairs produced in each distance bin and its subsequent evolution through interactions with the background medium. This is an inherently non-linear process, because the instabilities modify the beam distribution that drives them. Our phenomenological prescription should therefore be interpreted with caution.

\section{Discussion}
\label{sec:discussion}

The main strength of \texttt{grplinst} is that it treats the effect of plasma instabilities on gamma-ray signals at a purely phenomenological level, which is useful given the lack of consensus in the literature on the actual cooling rates and their dependence on beam and medium properties. This allows users to explore a wide range of models and parameters, and to propagate the associated uncertainty into observable predictions, without being tied to a specific theoretical framework or set of assumptions.

The approach adopted in the first version of \texttt{grplinst} was later adopted by another phenomenological implementation of plasma instabilities in the ELMAG code~\citep{blytt2020a}. In the current version, we kept this approach while further improving the modularity and flexibility of the code, allowing for a more systematic exploration of the parameter space and a clearer separation among the different ingredients entering the model. 

\citet{alvesbatista2019g} presented multiple prescriptions as bracketing models and highlighted that the topic is ``controversial'', explicitly noting that many \gls{IGMF} constraints had neglected instabilities and that non-linear damping and magnetic effects could change conclusions. Within that framework, the strongest cascade-quenching scenarios assume efficient dissipation, effectively taking $\tau_\text{pl}$ as close to the instability growth time as possible, precisely to explore a conservative lower envelope for the cascade flux.

Works relying on Monte Carlo and kinetic theory argued that nonlinear Landau damping and \gls{IGM} inhomogeneities could stabilise pair beams on timescales longer than inverse-Compton cooling, making plasma losses negligible for cascades in many cases~\citep{miniati2013a}. 
\gls{PIC} simulations suggested that for dilute relativistic beams, oblique modes can saturate early and deposit only a limited beam-energy fraction into the plasma~\citep{sironi2014a}. Linear analyses refined the conditions under which either kinetic or reactive treatments may be applied, and identified broad sets of unstable modes, demonstrating that the existence of an instability does not necessarily imply efficient beam cooling~\citep{bret2010a, schlickeiser2012a, schlickeiser2013a, chang2016a}.

These results emphasise that the existence of a linearly unstable mode does not, by itself, imply efficient beam cooling. The central uncertainties concern the non-linear saturation level, the feedback of the unstable waves on the beam, and the fraction of the wave energy that is ultimately transferred to the background plasma. Recent quasilinear calculations indicate that oblique modes may predominantly broaden the beam rather than remove a substantial fraction of its energy~\citep[see, e.g.,][]{perry2021a, alawashra2024a}. At the same time, additional plasma components can alter this conclusion: linear Landau damping caused by MeV cosmic-ray electrons may suppress oblique modes while enhancing the relative importance of quasi-parallel modes and their associated energy losses~\citep{alawashra2025b}. These results further motivate the phenomenological approach adopted here.

The generalisation to three dimensions is straightforward and seamless thanks to the modular design of the code. In fact, this can be achieved by simply running a standard CRPropa simulation of gamma-ray-induced cascades in a three-dimensional environment, and adding the \texttt{PlasmaInstability} module to the simulation pipeline. This allows us to explore the effects of phenomenological plasma cooling on cascade development in realistic three-dimensional magnetic-field configurations. However, this does not constitute a reliable self-consistent three-dimensional treatment of the beam--plasma system. In particular, weak tangled magnetic fields can broaden the momentum distribution of the pair beam and thereby reduce the growth rate of the electrostatic instability~\citep{alawashra2022a}. This coupling is not captured unless the prescribed cooling rate is made dependent on the local magnetic field and the angular distribution of the beam. A rigorous calculation would therefore require the self-consistent evolution of the spatial, momentum, and angular distributions of the beam, together with the spectrum and non-linear evolution of the plasma waves.

\section{Conclusion and Outlook}
\label{sec:outlook}

We have presented \texttt{grplinst}, a CRPropa-based plugin for exploring how the phenomenology of plasma-instabilities affect gamma-ray--induced electromagnetic cascades in the \gls{IGM}. 
Its architecture clearly separates the main ingredients -- beam, medium, geometry, and instability prescription -- making it straightforward to compare competing scenarios and to quantify how plasma-related assumptions propagate into observable gamma-ray signals.
The currently implemented prescriptions should, therefore, be interpreted as phenomenological models that encompass the theoretical uncertainty, rather than as definitive predictions of the energy loss by the beam.

It is important to stress that \texttt{grplinst}, as it currently stands, is not intended to be a plasma-physics code. Its purpose is not to simulate the plasma instabilities themselves in detail, but rather to model their phenomenological effects on cascade propagation. It therefore complements, rather than replaces, dedicated plasma-physics codes.

There are several natural directions for future development. The main one is to interface the code with plasma-physics tools to extract more realistic cooling rates and their dependence on beam and medium properties on the fly, as the beam propagates. By doing this iteratively, at small steps, one could capture the feedback between the beam and the plasma, which is expected to be important for understanding saturation and non-linear effects. This would be a significant step towards a more self-consistent treatment of plasma instabilities in cascade simulations, albeit at a much higher computational cost.

\medskip
The code presented here, \texttt{grplinst}, is available at \url{https://github.com/rafaelab/grplinst/}, together with documentation and examples. 
\bigskip

\section*{Conflict of Interest Statement}
The authors declare that the research was conducted in the absence of any commercial or financial relationships that could be construed as a potential conflict of interest.

\section*{Author Contributions}

Both authors contributed equally to the conceptualisation, code development, and writing of this paper.

\section*{Funding}
R.A.B. acknowledges support from the Agence Nationale de la Recherche (ANR), project ANR-23-CPJ1-0103-01.



\section*{Data Availability Statement}

The code described in this work is publicly available in the \texttt{grplinst} repository: \url{https://github.com/rafaelab/grplinst/}. 


\begin{thebibliography}{49}
\providecommand{\natexlab}[1]{#1}
\expandafter\ifx\csname urlstyle\endcsname\relax
  \providecommand{\doi}[1]{doi:\discretionary{}{}{}#1}\else
  \providecommand{\doi}{doi:\discretionary{}{}{}\begingroup \urlstyle{rm}\Url}\fi
\providecommand{\selectlanguage}[1]{\relax}
\providecommand{\bibAnnoteFile}[1]{%
  \IfFileExists{#1}{\begin{quotation}\noindent\textsc{Key:} #1\\
  \textsc{Annotation:}\ \input{#1}\end{quotation}}{}}
\providecommand{\bibAnnote}[2]{%
  \begin{quotation}\noindent\textsc{Key:} #1\\
  \textsc{Annotation:}\ #2\end{quotation}}

\bibitem[{Alawashra and Pohl(2022)}]{alawashra2022a}
Alawashra, M. and Pohl, M. (2022).
\newblock Suppression of the TeV pair-beam-plasma instability by a tangled weak intergalactic magnetic field.
\newblock \emph{The Astrophysical Journal} 929, 67.
\newblock \doi{10.3847/1538-4357/ac5a4b}
\input{alawashra2022a}

\bibitem[{Alawashra and Pohl(2024)}]{alawashra2024a}
Alawashra, M. and Pohl, M. (2024).
\newblock Nonlinear feedback of the electrostatic instability on the blazar-induced pair beam and GeV cascade.
\newblock \emph{The Astrophysical Journal} 964, 82.
\newblock \doi{10.3847/1538-4357/ad24ea}
\input{alawashra2024a}

\bibitem[{Alawashra et~al.(2025)Alawashra, Yang, Hirata, Long, and Pohl}]{alawashra2025b}
Alawashra, M., Yang, Y., Hirata, C.~M., Long, H., and Pohl, M. (2025).
\newblock MeV cosmic-ray electrons modify the TeC pair-beam plasma instability.
\newblock \emph{The Astrophysical Journal} 989, 37.
\newblock \doi{10.3847/1538-4357/adec9e}
\input{alawashra2025b}

\bibitem[{Alves~Batista et~al.(2022)Alves~Batista, Becker~Tjus, D\"orner, Dundovic, Eichmann, Frie et~al.}]{alvesbatista2022a}
Alves~Batista, R., Becker~Tjus, J., D\"orner, J., Dundovic, A., Eichmann, B., Frie, A., et~al. (2022).
\newblock CRPropa 3.2 -- an advanced framework for high-energy particle propagation in extragalactic and galactic spaces.
\newblock \emph{Journal of Cosmology and Astroparticle Physics} 09, 035.
\newblock \doi{10.1088/1475-7516/2022/09/035}
\input{alvesbatista2022a}

\bibitem[{Alves~Batista et~al.(2016)Alves~Batista, Dundovic, Erdmann, Kampert, Kuempel, M\"uller et~al.}]{alvesbatista2016a}
Alves~Batista, R., Dundovic, A., Erdmann, M., Kampert, K.-H., Kuempel, D., M\"uller, G., et~al. (2016).
\newblock CRPropa 3 - a public astrophysical simulation framework for propagating extraterrestrial ultra-high energy particles.
\newblock \emph{Journal of Cosmology and Astroparticle Physics} 5, 038.
\newblock \doi{10.1088/1475-7516/2016/05/038}
\input{alvesbatista2016a}

\bibitem[{Alves~Batista and Saveliev(2021)}]{alvesbatista2021a}
Alves~Batista, R. and Saveliev, A. (2021).
\newblock The gamma-ray window to intergalactic magnetism.
\newblock \emph{Universe} 7, 223.
\newblock \doi{10.3390/universe7070223}
\input{alvesbatista2021a}

\bibitem[{Alves~Batista et~al.(2019)Alves~Batista, Saveliev, and de~Gouveia Dal~Pino}]{alvesbatista2019g}
Alves~Batista, R., Saveliev, A., and de~Gouveia Dal~Pino, E.~M. (2019).
\newblock The impact of plasma instabilities on the spectra of TeV blazars.
\newblock \emph{Monthly Notices of the Royal Astronomical Society} 489, arXiv:1904.13345.
\newblock \doi{10.1093/mnras/stz2389}
\input{alvesbatista2019g}

\bibitem[{Beazley(1996)}]{beazley1996a}
Beazley, D.~M. (1996).
\newblock Swig: an easy to use tool for integrating scripting languages with C and C++.
\newblock In \emph{Proceedings of the 4th Conference on USENIX Tcl/Tk Workshop, 1996 - Volume 4} (USA: USENIX Association), TCLTK'96, 15
\input{beazley1996a}

\bibitem[{Blytt et~al.(2020)Blytt, Kachelrie\ss{}, and Ostapchenko}]{blytt2020a}
Blytt, M., Kachelrie\ss{}, M., and Ostapchenko, S. (2020).
\newblock ELMAG 3.01: A three-dimensional monte carlo simulation of electromagnetic cascades on the extragalactic background light and in magnetic fields.
\newblock \emph{Computer Physics Communications} 252, 107163.
\newblock \doi{10.1016/j.cpc.2020.107163}
\input{blytt2020a}

\bibitem[{Bonnoli et~al.(2015)Bonnoli, Tavecchio, Ghisellini, and Sbarrato}]{bonnoli2015a}
Bonnoli, G., Tavecchio, F., Ghisellini, G., and Sbarrato, T. (2015).
\newblock An emerging population of BL Lacs with extreme properties: towards a class of ebl and cosmic magnetic field probes?
\newblock \emph{Monthly Notices of the Royal Astronomical Society} 451, 611--621.
\newblock \doi{10.1093/mnras/stv953}
\input{bonnoli2015a}

\bibitem[{Bret et~al.(2010)Bret, Gremillet, and Dieckmann}]{bret2010a}
Bret, A., Gremillet, L., and Dieckmann, M.~E. (2010).
\newblock Multidimensional electron beam-plasma instabilities in the relativistic regime.
\newblock \emph{Physics of Plasmas} 17, 120501.
\newblock \doi{10.1063/1.3514586}
\input{bret2010a}

\bibitem[{Broderick et~al.(2012)Broderick, Chang, and Pfrommer}]{broderick2012a}
Broderick, A.~E., Chang, P., and Pfrommer, C. (2012).
\newblock The cosmological impact of luminous TeV blazars. i. implications of plasma instabilities for the intergalactic magnetic field and extragalactic gamma-ray background.
\newblock \emph{The Astrophysical Journal} 752, 22.
\newblock \doi{10.1088/0004-637X/752/1/22}
\input{broderick2012a}

\bibitem[{Broderick et~al.(2014)Broderick, Pfrommer, Puchwein, and Chang}]{broderick2014a}
Broderick, A.~E., Pfrommer, C., Puchwein, E., and Chang, P. (2014).
\newblock Implications of plasma beam instabilities for the statistics of the fermi hard gamma-ray blazars and the origin of the extragalactic gamma-ray background.
\newblock \emph{The Astrophysical Journal} 790, 137.
\newblock \doi{10.1088/0004-637X/790/2/137}
\input{broderick2014a}

\bibitem[{Celotti and Ghisellini(2008)}]{celotti2008a}
Celotti, A. and Ghisellini, G. (2008).
\newblock The power of blazar jets.
\newblock \emph{Monthly Notices of the Royal Astronomical Society} 385, 283--300.
\newblock \doi{10.1111/j.1365-2966.2007.12758.x}
\input{celotti2008a}

\bibitem[{Chang et~al.(2012)Chang, Broderick, and Pfrommer}]{chang2012a}
Chang, P., Broderick, A.~E., and Pfrommer, C. (2012).
\newblock The cosmological impact of luminous TeV blazars. II. rewriting the thermal history of the intergalactic medium.
\newblock \emph{The Astrophysical Journal} 752, 23.
\newblock \doi{10.1088/0004-637X/752/1/23}
\input{chang2012a}

\bibitem[{Chang et~al.(2014)Chang, Broderick, Pfrommer, Puchwein, Lamberts, and Shalaby}]{chang2014a}
Chang, P., Broderick, A.~E., Pfrommer, C., Puchwein, E., Lamberts, A., and Shalaby, M. (2014).
\newblock The effect of nonlinear landau damping on ultrarelativistic beam plasma instabilities.
\newblock \emph{The Astrophysical Journal} 797, 110.
\newblock \doi{10.1088/0004-637X/797/2/110}
\input{chang2014a}

\bibitem[{Chang et~al.(2016)Chang, Broderick, Pfrommer, Puchwein, Lamberts, Shalaby et~al.}]{chang2016a}
Chang, P., Broderick, A.~E., Pfrommer, C., Puchwein, E., Lamberts, A., Shalaby, M., et~al. (2016).
\newblock The linear instability of dilute ultrarelativistic $e^{\pm}$ pair beams.
\newblock \emph{The Astrophysical Journal} 833, 118.
\newblock \doi{10.3847/1538-4357/833/1/118}
\input{chang2016a}

\bibitem[{Chang et~al.(2019)Chang, Arsioli, Giommi, Padovani, and Brandt}]{chang2019a}
Chang, Y.~L., Arsioli, B., Giommi, P., Padovani, P., and Brandt, C.~H. (2019).
\newblock The 3HSP catalogue of extreme and high-synchrotron peaked blazars.
\newblock \emph{Astronomy and Astrophysics} 632, A77.
\newblock \doi{10.1051/0004-6361/201834526}
\input{chang2019a}

\bibitem[{Chen et~al.(2018)Chen, Errando, and Buckley}]{chen2018a}
Chen, W., Errando, M., and Buckley, J. (2018).
\newblock Novel search for TeV-initiated pair cascades in the intergalactic medium.
\newblock In \emph{42nd COSPAR Scientific Assembly}. vol.~42, E1.14--26--18.
\newblock \doi{10.48550/arXiv.1811.05774}
\input{chen2018a}

\bibitem[{Collaboration(2023)}]{magic2023a}
MAGIC Collaboration (2023).
\newblock A lower bound on intergalactic magnetic fields from time variability of 1ES\textasciitilde{}0229+200 from magic and fermi/lat observations.
\newblock \emph{Astronomy and Astrophysics} 670, A145.
\newblock \doi{10.1051/0004-6361/202244126}
\input{magic2023a}

\bibitem[{Costamante(2020)}]{costamante2020a}
Costamante, L. (2020).
\newblock TeV-peaked candidate BL Lac objects.
\newblock \emph{Monthly Notices of the Royal Astronomical Society} 491, 2771--2778.
\newblock \doi{10.1093/mnras/stz3018}
\input{costamante2020a}

\bibitem[{Gould and Schr\'eder(1967)}]{gould1967a}
Gould, R.~J. and Schr\'eder, G.~P. (1967).
\newblock Opacity of the universe to high-energy photons.
\newblock \emph{Physical Review} 155, 1408--1411.
\newblock \doi{10.1103/PhysRev.155.1408}
\input{gould1967a}

\bibitem[{Grognard(1975)}]{grognard1975a}
Grognard, R. J.~M. (1975).
\newblock Deficiencies of the asymptotic solutions commonly found in the quasilinear relaxation theory.
\newblock \emph{Australian Journal of Physics} 28, 731.
\newblock \doi{10.1071/PH750731}
\input{grognard1975a}

\bibitem[{Harris et~al.(2020)Harris, Millman, van~der Walt, Gommers, Virtanen, Cournapeau et~al.}]{harris2020a}
Harris, C.~R., Millman, K.~J., van~der Walt, S.~J., Gommers, R., Virtanen, P., Cournapeau, D., et~al. (2020).
\newblock Array programming with numpy.
\newblock \emph{Nature} 585, 357--362.
\newblock \doi{10.1038/s41586-020-2649-2}
\input{harris2020a}

\bibitem[{Kempf et~al.(2016)Kempf, Kilian, and Spanier}]{kempf2016a}
Kempf, A., Kilian, P., and Spanier, F. (2016).
\newblock Energy loss in intergalactic pair beams: Particle-in-cell simulation.
\newblock \emph{Astronomy and Astrophysics} 585, A132.
\newblock \doi{10.1051/0004-6361/201527521}
\input{kempf2016a}

\bibitem[{Lamberts et~al.(2015)Lamberts, Chang, Pfrommer, Puchwein, Broderick, and Shalaby}]{lamberts2015a}
Lamberts, A., Chang, P., Pfrommer, C., Puchwein, E., Broderick, A.~E., and Shalaby, M. (2015).
\newblock Patchy blazar heating: Diversifying the thermal history of the intergalactic medium.
\newblock \emph{The Astrophysical Journal} 811, 19.
\newblock \doi{10.1088/0004-637X/811/1/19}
\input{lamberts2015a}

\bibitem[{McQuinn(2016)}]{mcquinn2016a}
McQuinn, M. (2016).
\newblock The evolution of the intergalactic medium.
\newblock \emph{Annual Review of Astronomy and Astrophysics} 54, 313--362.
\newblock \doi{10.1146/annurev-astro-082214-122355}
\input{mcquinn2016a}

\bibitem[{Meiksin(2009)}]{meiksin2009a}
Meiksin, A.~A. (2009).
\newblock The physics of the intergalactic medium.
\newblock \emph{Reviews of Modern Physics} 81, 1405--1469.
\newblock \doi{10.1103/RevModPhys.81.1405}
\input{meiksin2009a}

\bibitem[{Miniati and Elyiv(2013)}]{miniati2013a}
Miniati, F. and Elyiv, A. (2013).
\newblock Relaxation of blazar-induced pair beams in cosmic voids.
\newblock \emph{The Astrophysical Journal} 770, 54.
\newblock \doi{10.1088/0004-637X/770/1/54}
\input{miniati2013a}

\bibitem[{Neronov and Semikoz(2009)}]{neronov2009a}
Neronov, A. and Semikoz, D.~V. (2009).
\newblock Sensitivity of {$\gamma$}-ray telescopes for detection of magnetic fields in the intergalactic medium.
\newblock \emph{Physical Review D} 80, 123012.
\newblock \doi{10.1103/PhysRevD.80.123012}
\input{neronov2009a}

\bibitem[{Neronov and Vovk(2010)}]{neronov2010a}
Neronov, A. and Vovk, I. (2010).
\newblock Evidence for strong extragalactic magnetic fields from fermi observations of TeV blazars.
\newblock \emph{Science} 328, 73.
\newblock \doi{10.1126/science.1184192}
\input{neronov2010a}

\bibitem[{Pavan et~al.(2011)Pavan, Yoon, and Umeda}]{pavan2011a}
Pavan, J., Yoon, P.~H., and Umeda, T. (2011).
\newblock Quasilinear theory and simulation of buneman instability.
\newblock \emph{Physics of Plasmas} 18, 042307.
\newblock \doi{10.1063/1.3574359}
\input{pavan2011a}

\bibitem[{Perry and Lyubarsky(2021)}]{perry2021a}
Perry, R. and Lyubarsky, Y. (2021).
\newblock The role of resonant plasma instabilities in the evolution of blazar-induced pair beams.
\newblock \emph{Monthly Notices of the Royal Astronomical Society} 503, 2215--2228.
\newblock \doi{10.1093/mnras/stab324}
\input{perry2021a}

\bibitem[{Plaga(1995)}]{plaga1995a}
Plaga, R. (1995).
\newblock Detecting intergalactic magnetic fields using time delays in pulses of gamma-rays.
\newblock \emph{Nature} 374, 430
\input{plaga1995a}

\bibitem[{Pohl et~al.(2020)Pohl, Hoshino, and Niemiec}]{pohl2020a}
Pohl, M., Hoshino, M., and Niemiec, J. (2020).
\newblock PIC simulation methods for cosmic radiation and plasma instabilities.
\newblock \emph{Progress in Particle and Nuclear Physics} 111, 103751.
\newblock \doi{10.1016/j.ppnp.2019.103751}
\input{pohl2020a}

\bibitem[{Puchwein et~al.(2012)Puchwein, Pfrommer, Springel, Broderick, and Chang}]{puchwein2012a}
Puchwein, E., Pfrommer, C., Springel, V., Broderick, A.~E., and Chang, P. (2012).
\newblock The Lyman $\alpha$ forest in a blazar-heated universe.
\newblock \emph{Monthly Notices of the Royal Astronomical Society} 423, 149--164.
\newblock \doi{10.1111/j.1365-2966.2012.20738.x}
\input{puchwein2012a}

\bibitem[{Rafighi et~al.(2017)Rafighi, Vafin, Pohl, and Niemiec}]{rafighi2017a}
Rafighi, I., Vafin, S., Pohl, M., and Niemiec, J. (2017).
\newblock Plasma effects on relativistic pair beams from TeV blazars: PIC simulations and analytical predictions.
\newblock \emph{Astronomy and Astrophysics} 607, A112.
\newblock \doi{10.1051/0004-6361/201731127}
\input{rafighi2017a}

\bibitem[{Schlickeiser et~al.(2012)Schlickeiser, Ibscher, and Supsar}]{schlickeiser2012a}
Schlickeiser, R., Ibscher, D., and Supsar, M. (2012).
\newblock Plasma effects on fast pair beams in cosmic voids.
\newblock \emph{The Astrophysical Journal} 758, 102.
\newblock \doi{10.1088/0004-637X/758/2/102}
\input{schlickeiser2012a}

\bibitem[{Schlickeiser et~al.(2013)Schlickeiser, Krakau, and Supsar}]{schlickeiser2013a}
Schlickeiser, R., Krakau, S., and Supsar, M. (2013).
\newblock Plasma effects on fast pair beams. ii. reactive versus kinetic instability of parallel electrostatic waves.
\newblock \emph{The Astrophysical Journal} 777, 49.
\newblock \doi{10.1088/0004-637X/777/1/49}
\input{schlickeiser2013a}

\bibitem[{Shalaby et~al.(2018)Shalaby, Broderick, Chang, Pfrommer, Lamberts, and Puchwein}]{shalaby2018a}
Shalaby, M., Broderick, A.~E., Chang, P., Pfrommer, C., Lamberts, A., and Puchwein, E. (2018).
\newblock Growth of beam-plasma instabilities in the presence of background inhomogeneity.
\newblock \emph{The Astrophysical Journal} 859, 45.
\newblock \doi{10.3847/1538-4357/aabe92}
\input{shalaby2018a}

\bibitem[{Shalaby et~al.(2020)Shalaby, Broderick, Chang, Pfrommer, Puchwein, and Lamberts}]{shalaby2020a}
Shalaby, M., Broderick, A.~E., Chang, P., Pfrommer, C., Puchwein, E., and Lamberts, A. (2020).
\newblock The growth of the longitudinal beam-plasma instability in the presence of an inhomogeneous background.
\newblock \emph{Journal of Plasma Physics} 86, 535860201.
\newblock \doi{10.1017/S0022377820000215}
\input{shalaby2020a}

\bibitem[{Sikora(2016)}]{sikora2016a}
Sikora, M. (2016).
\newblock Powers and magnetization of blazar jets.
\newblock \emph{Galaxies} 4, 12.
\newblock \doi{10.3390/galaxies4030012}
\input{sikora2016a}

\bibitem[{Sironi and Giannios(2014)}]{sironi2014a}
Sironi, L. and Giannios, D. (2014).
\newblock Relativistic pair beams from TeV blazars: A source of reprocessed GeV emission rather than intergalactic heating.
\newblock \emph{The Astrophysical Journal} 787, 49.
\newblock \doi{10.1088/0004-637X/787/1/49}
\input{sironi2014a}

\bibitem[{Sol and Zech(2022)}]{sol2022a}
Sol, H. and Zech, A. (2022).
\newblock Blazars at very high energies: Emission modelling.
\newblock \emph{Galaxies} 10, 105.
\newblock \doi{10.3390/galaxies10060105}
\input{sol2022a}

\bibitem[{Supsar and Schlickeiser(2014)}]{supsar2014a}
Supsar, M. and Schlickeiser, R. (2014).
\newblock Plasma effects on fast pair beams. iii. oblique electrostatic growth rates for perpendicular maxwellian pair beams.
\newblock \emph{The Astrophysical Journal} 783, 96.
\newblock \doi{10.1088/0004-637X/783/2/96}
\input{supsar2014a}

\bibitem[{Tavecchio et~al.(2010)Tavecchio, Ghisellini, Foschini, Bonnoli, Ghirlanda, and Coppi}]{tavecchio2010a}
Tavecchio, F., Ghisellini, G., Foschini, L., Bonnoli, G., Ghirlanda, G., and Coppi, P. (2010).
\newblock The intergalactic magnetic field constrained by fermi/large area telescope observations of the TeV blazar 1ES0229+200.
\newblock \emph{Monthly Notices of the Royal Astronomical Society} 406, L70--L74.
\newblock \doi{10.1111/j.1745-3933.2010.00884.x}
\input{tavecchio2010a}

\bibitem[{Vafin et~al.(2019)Vafin, Deka, Pohl, and Bohdan}]{vafin2019a}
Vafin, S., Deka, P.~J., Pohl, M., and Bohdan, A. (2019).
\newblock Revisit of nonlinear landau damping for electrostatic instability driven by blazar-induced pair beams.
\newblock \emph{The Astrophysical Journal} 873, 10.
\newblock \doi{10.3847/1538-4357/ab017b}
\input{vafin2019a}

\bibitem[{Vafin et~al.(2018)Vafin, Rafighi, Pohl, and Niemiec}]{vafin2018a}
Vafin, S., Rafighi, I., Pohl, M., and Niemiec, J. (2018).
\newblock The electrostatic instability for realistic pair distributions in blazar/ebl cascades.
\newblock \emph{The Astrophysical Journal} 857, 43.
\newblock \doi{10.3847/1538-4357/aab552}
\input{vafin2018a}

\bibitem[{Van Der~Walt et~al.(2011)Van Der~Walt, Colbert, and Varoquaux}]{vanderwalt2011a}
Van Der~Walt, S., Colbert, S.~C., and Varoquaux, G. (2011).
\newblock The numpy array: A structure for efficient numerical computation.
\newblock \emph{Computing in Science and Engineering} 13, 22--30.
\newblock \doi{10.1109/MCSE.2011.37}
\input{vanderwalt2011a}

\end{thebibliography}

\end{document}